\documentclass[twocolumn,superscriptaddress]{revtex4}  

\usepackage{graphicx}  
\usepackage{dcolumn}   
\usepackage{bm}        
\usepackage{amssymb}   

\usepackage{amsmath}
\usepackage{amssymb}
\usepackage{natbib}
\usepackage{bm}

\begin{document}

\title{All-electric all-semiconductor spin field effect transistors}


\author{Pojen Chuang,$^{1\dagger}$ Sheng-Chin Ho,$^{1\dagger}$ L. W. Smith,$^{2}$ F. Sfigakis,$^{2}$ M. Pepper,$^{3}$ Chin-Hung Chen,$^{1}$ Ju-Chun Fan,$^{1}$ J. P. Griffiths,$^{2}$ I. Farrer,$^{2}$ H. E. Beere,$^{2}$ G. A. C. Jones,$^{2}$ D. A. Ritchie,$^{2}$ \& T.-M. Chen$^{1\ast}$\\
\vspace*{2mm}
\normalsize{$^{1}$\it{Department of Physics, National Cheng Kung University, Tainan 701, Taiwan}}\\
\normalsize{$^{2}$\it{Cavendish Laboratory, J J Thomson Avenue, Cambridge CB3 0HE, United Kingdom}}\\
\normalsize{$^{3}$\it{Department of Electronic and Electrical Engineering, University College London, London WC1E 7JE, United Kingdom}}\\
\normalsize{$^\dagger$These authors contributed equally to the work.}\\
\normalsize{$^\ast$e-mail: tmchen@mail.ncku.edu.tw}
}

\begin{abstract}
The spin field effect transistor envisioned by Datta and Das\cite{datta_apl90} opens a gateway to spin information processing\cite{zutic_rmp04,awschalom_np07}. Although the coherent manipulation of electron spins in semiconductors is now possible\cite{crooker_science05,appelbaum_nature07,lou_np07,koo_science09}, the realization of a functional spin field effect transistor for information processing has yet to be achieved, owing to several fundamental challenges such as the low spin-injection efficiency due to resistance mismatch\cite{schmidt_prb00}, spin relaxation, and the spread of spin precession angles. Alternative spin transistor designs have therefore been proposed\cite{wunderlich_science10,betthausen_science12}, but these differ from the field effect transistor concept and require the use of optical or magnetic elements, which pose difficulties for the incorporation into integrated circuits. Here, we present an all-electric and all-semiconductor spin field effect transistor, in which these obstacles are overcome by employing two quantum point contacts as spin injectors and detectors. Distinct engineering architectures of spin-orbit coupling are exploited for the quantum point contacts and the central semiconductor channel to achieve complete control of the electron spins---spin injection, manipulation, and detection---in a purely electrical manner. Such a device is compatible with large-scale integration and hold promise for future spintronic devices for information processing.
\end{abstract}

\maketitle

Spin-orbit (SO) coupling---the interaction between a particle's spin and its motion---can be appreciated in the framework of an effective magnetic field $\textbf{B}^{\text{SO}}$, which acts on charged particles when they move in an electric field $\textbf{E}$ and is described by $\textbf{B}^{\text{SO}} = - (\hbar/mc^2) (\textbf{k} \times \textbf{E})$, where $\hbar$ is Planck's constant divided by $2\pi$,  $c$ is the speed of light, $\textbf{k}$ is the particle's wavevector, and $m$ is its mass. In semiconductor heterostructures, the electric field which gives rise to $\textbf{B}^{\text{SO}}$ can be created by breaking the structural inversion symmetry in the material, namely, the Rashba SO coupling\cite{rashba_60,bychovand_jpc84}. Moreover, this electric field can easily be varied using metallic gates\cite{nitta_prl97,koga_prl02}, thus controlling $\textbf{B}^{\text{SO}}$. Such an effective magnetic field creates a link between the magnetic moment of the particle (spin) and the electric field acting upon it, offering a route for fast and coherent electrical control of spin states. While the SO coupling has been utilized for spin manipulation, approaches to spin injection and detection still rely on ferromagnetic and/or optical components, and the demonstration of an all-electric spin transistor device has remained elusive.

\begin{figure}
\begin{center}
\includegraphics[width=0.95\columnwidth]{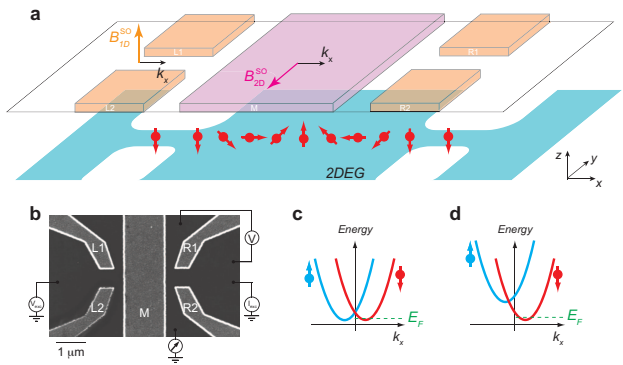}
\end{center}
\caption{\small \textbf{All-electric and all-semiconductor spin FET.} \textbf{a},\textbf{b}, Schematic (a) and electron-microscope image (b) of an all-electric spin FET device. The left (right) QPC, consisting of a pair of split gates L1 and L2 (R1 and R2), acts as a spin injector (detector) when the split gates are asymmetrically biased to generate a lateral inversion asymmetry and consequently a spin-orbit (SO) effective magnetic field $B^{\text{SO}}_{\text{1D}}$. The injected spins, polarized along the $z$ axis, move ballistically and precess about the $y$ axis in the region between the two QPCs. The precession originates from a distinct SO effective field $B^{\text{SO}}_{\text{2D}}$ which is defined and controlled by the structural inversion asymmetry of the 2DEG channel and the middle gate (M) voltage. Electrons can pass through the QPC detector if their spin rotates to be parallel to the polarization direction, and cannot pass if their spin is anti-parallel. \textbf{c}, The dispersion relation of 1D subbands with SO coupling, where the spin-down (red) and spin-up (blue) subbands are laterally shifted. If the Fermi energy $E_F$ (green dashed line) lies below the crossing point between two spin-polarized subbands, only one spin-species is present in either the right- ($+k_x$) or left- ($-k_x$) moving directions. \textbf{d}, Electron-electron interactions shift the spin-up and down subbands vertically and enhance the spin-orbit induced spin splitting.}
\end{figure}

Figure~1 illustrates our proposed spin field effect transistor (FET) and its operating principle. An InGaAs heterostructure (see Methods Summary), one of the strong contenders to replace Si in future generations of large-scale integrated circuits (see \textit{International Technology Roadmap for Semiconductors}; http://public.itrs.net), is used to provide a two-dimensional electron gas (2DEG) channel for ballistic electron transport under a metallic middle gate and between two gate-defined quantum point contacts (QPCs). The QPCs are narrow and short one-dimensional (1D) constrictions, usually formed by applying voltages to split gates patterned on the surface of a semiconductor heterostructure. Although the geometry is extremely simple, the QPC contains rich physics\cite{thomas_prl96,bauer_nature13,iqbal_nature13} and has been suggested to generate a completely spin-polarized current due to SO coupling and/or electron-electron interaction\cite{debray_naturenano09,wan_prb09,quay_naturephys10,nowak_apl13,chen_apl08,chen_prl12}.

In this all-electric spin FET, the left (right) QPC acts as a spin injector (detector) with nearly $100\%$ efficiency. To utilize the QPCs as spin injectors/detectors, we set a difference between the voltages on either side of the split gate (i.e. $V_{\text{L1}} - V_{\text{L2}} \neq 0$ where $V_{\text{L1}}$ and $V_{\text{L2}}$ are the voltages applied respectively to the gate L1 and L2 in Fig.~1a,~b) to generate a lateral inversion asymmetry and consequently a lateral SO effective magnetic field, $\textbf{B}^{\text{SO}}_{\text{1D}}$, on electrons moving within the 1D constriction. The orientation of $\textbf{B}^{\text{SO}}_{\text{1D}}$ is along the $z$ axis, perpendicular to the lateral electric field and the electron momentum direction. Such a lateral SO coupling lifts the spin degeneracy and results in two spin-polarized 1D subbands shifted in wavevector as shown in Fig.~1c. In the case where the Fermi energy $E_F$ is tuned below the crossing point between two spin-polarized subbands, the left- and right-moving 1D electrons are both fully spin-polarized\cite{moroz_prb99} in the positive and negative $z$-direction, respectively (hereafter, we refer to these subbands as the spin-up and spin-down states), thereby allowing the QPC to act as a spin injector/detector. Recent studies\cite{debray_naturenano09,wan_prb09} have further suggested that this lateral SO-induced spin splitting could be greatly enhanced by the strong electron-electron interaction in 1D systems (Fig.~1d), making the QPC spin injector/detector more reliable (see Supplementary Section 1). This method of spin injection circumvents many of the technical problems faced by ferromagnetic or optical alternatives (such as low spin-injection efficiency\cite{schmidt_prb00} and scalability), and is compatible with the current manufacturing technology of FETs.

The spins supplied from the QPC injector remain ballistic and experience a SO effective magnetic field, $\textbf{B}^{\text{SO}}_{\text{2D}}$, in the 2DEG channel due to the structural inversion asymmetry of the quantum well, which can be further controlled by changing the voltage applied to the middle gate ($V_{\text{M}}$). In this transport channel the orientation of $\textbf{B}^{\text{SO}}_{\text{2D}}$ is parallel to the $y$ axis, and therefore perpendicular to the SO field $\textbf{B}^{\text{SO}}_{\text{1D}}$ in the QPC injector. This causes the injected spins to precess during transport between the QPCs (Fig. 1a). By modifying the gate voltage $V_{\text{M}}$ to vary $B^{\text{SO}}_{\text{2D}}$, one can control the spin orientation of electrons travelling along the channel. The charge current is therefore modulated by the spin precession angle: electrons can pass through the QPC detector if their spin rotates to become parallel to the polarization direction, and cannot if their spin is anti-parallel. This gives rise to an oscillatory on/off switching with respect to gate voltage $V_{\text{M}}$. 

\begin{figure}
\begin{center}
\includegraphics[width=0.95\columnwidth]{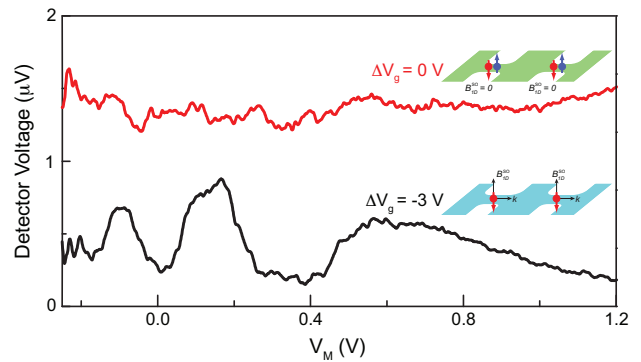}
\end{center}
\caption{\small \textbf{Oscillating on/off switch of the spin FET.} Detector voltage as a function of gate voltage $V_{\text{M}}$ (which controls the spin precession frequency) measured at $T = 30$~mK and $G_{\text{QPC}}=0.3G_0$ (where $G_0=2e^2/h$). The oscillating current modulation occurs when a voltage difference $\Delta V_g = V_{\text{L1}} - V_{\text{L2}} = V_{\text{R1}} - V_{\text{R2}} = -3$~V is applied to the QPCs (black trace). The lateral asymmetry of the QPC confinement potential results in a lateral SO effective field $\textbf{B}^{\text{SO}}_{\text{1D}}$ on electrons moving within the 1D channel, and hence the QPCs acts as spin injectors/detectors when operated near threshold (see bottom inset, schematic of the spin FET). The oscillation disappears at $\Delta V_g=0$ (red trace), where  the lateral SO effective field is absent, $\textbf{B}^{\text{SO}}_{\text{1D}}=0$, and both spin species can pass through the QPCs (see top inset). Data are vertically offset by $1$~$\mu$V for clarity. }
\end{figure}

We demonstrate the operation of our spin FET in Fig.~2. Here, in order to simultaneously measure the on/off switching functionality and have precise control of the conductance of the QPCs, we configured the QPC detector as a voltage probe and measured the voltage across it. This voltage corresponds to the current flowing directly from the injector into the detector (see Methods Summary), i.e., the switching current in the spin FET. The conductance values of both QPCs are just above the threshold for conduction set at $G_{\text{QPC}}=0.3 \times 2e^2/h$ (where $e$ is the electron charge), at which the Fermi level is slightly above the very bottom of the spin-polarized 1D subbands to generate a spin polarized current in the presence of $B^{\text{SO}}_{\text{1D}}$. When both QPCs are brought into the spin-polarized state by electrically introducing a lateral inversion asymmetry (black trace; $\Delta V_g = V_{\text{L1}} - V_{\text{L2}} = V_{\text{R1}} - V_{\text{R2}} = -3$~V where $V_{\text{R1}}$ and $V_{\text{R2}}$ are the split gate voltages), an oscillatory on/off switching with variation as high as $500\%$ is observed as a function of $V_{\text{M}}$. Such a large oscillating change in the conductance modulation (due to $\textbf{B}^{\text{SO}}$ and spin precession) is about $100,000$ times greater than that observed in a conventional 2D spin FET design\cite{koo_science09} which suffers from low signal levels as a result of the limited spin-injection efficiency, the short spin lifetime, and the spread of spin precession angles.

The voltage oscillation disappears when the lateral inversion asymmetry is removed from the QPCs by setting $\Delta V_g = 0$ (red trace in Fig. 2). Spins injected from the QPC are no longer polarized along the $z$ axis as $B^{\text{SO}}_{\text{1D}}=0$, and thus no oscillations in current are detected. It is worth noting that the experimental results presented here, in addition to showing the realization of spin FETs, provide the first direct evidence of spin polarization of QPCs at zero external magnetic field.

\begin{figure}
\begin{center}
\includegraphics[width=0.95\columnwidth]{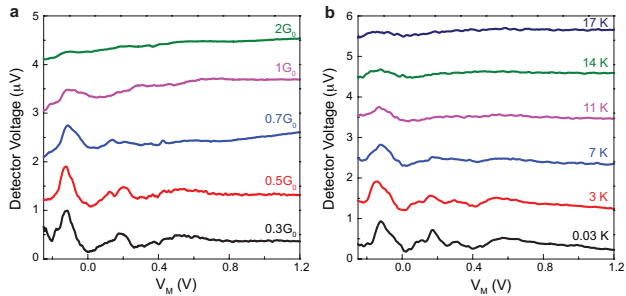}
\end{center}
\caption{\small \textbf{Influence of QPC conductance and temperature on the operation of spin FETs.} \textbf{a}, Detector voltage as a function of $V_{\text{M}}$ at various $G_{\text{QPC}}$ values, ranging from $0.3G_0$ to $2G_0$, while $T$ is fixed at $0.03$~K. Data are vertically offset by $1$~$\mu$V for clarity. \textbf{b}, Same as \textbf{a} for various temperatures ranging from $0.03$ to $17$~K, for $G_{\text{QPC}} = 0.3G_0$.}
\end{figure}

Figure~3a shows the oscillating voltages when the injector and detector QPCs are set at various conductance values. In a simple model of 1D transport with SO coupling (Fig.~1c), the right-moving electrons (with $+k_x$ wavevectors) are fully spin-polarized at low conductance values when only the lowest spin-down subband is occupied. With increasing $G_{\text{QPC}}$, the 1D subbands of both spin species become populated by electrons and the spin polarization decreases. Fig.~3a shows that the oscillation amplitude decreases with increasing $G_{\text{QPC}}$, which is consistent with this model.

The influence of temperature on the oscillating voltage was also investigated (Fig.~3b). Since momentum scattering plays a key role in randomizing the spin precession\cite{dyakonov_jetp71,dyakonov_sps86,elliott_pr54}, in a collision-free regime the spin relaxation may be negligible. The use of QPCs in the spin FET device allows only the ballistic transport electrons that directly moves from the injector to the collector to contribute to the signal, thereby implying that the observed decrease of the oscillation amplitude mainly results from the thermal reduction of the QPC polarization efficiency rather than the spin relaxation during transport. It suggests that much higher working temperature of the spin FET could be achieved in the presence of a larger 1D spin splitting, perhaps using wet-etched QPCs\cite{debray_naturenano09} or InAs nanowires.

\begin{figure}
\begin{center}
\includegraphics[width=0.95\columnwidth]{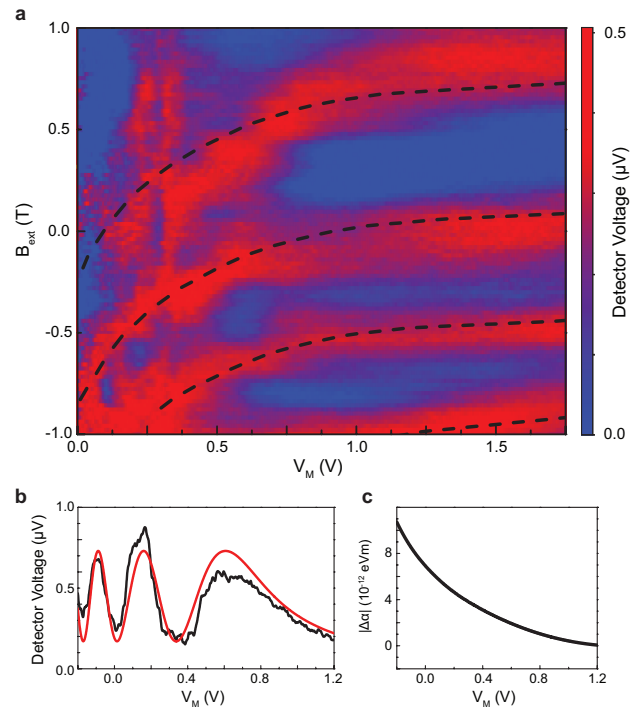}
\end{center}
\caption{\small \textbf{Simultaneous electrical and magnetic control of spin precession.} \textbf{a}, The spectrum of spin precession angle as a function of electrical gate voltage $V_{\text{M}}$ and magnetic field $B_{\text{ext}}$, obtained in a different cooldown to data in Fig. 2 and 3. The dashed lines show the calculated positions of oscillation peaks (i.e., the spin precession angle $\theta = 2n\pi$), in good quantitative agreement with the experiments. \textbf{b}, Experimental data of the oscillating voltage at $B_{\text{ext}} = 0$ (black trace) and its fit using Eq. 1 in cosine form (red trace). \textbf{c}, The SO coupling variation $|\Delta \alpha|$ as a function of gate voltage $V_{\text{M}}$, obtained from the fit in \textbf{b}. Note that the analysis of spin precession with respect to $V_{\text{M}}$ can only provide the absolute value of $\Delta \alpha$. However, the interplay between the external field and the Rashba SO field on spin precession in \textbf{a} can be used to verify the direction of the Rashba SO field, showing $\alpha (V_{\text{M}}) = \alpha_\text{b} + |\Delta \alpha (V_{\text{M}})|$ is a negative value and decreases with creasing $V_{\text{M}}$, where $\alpha_\text{b}$ is a baseline value of the Rashba SO coupling constant.}
\end{figure}

Finally, we demonstrate simultaneous electrical and magnetic control of spin precession. Earlier studies have shown that the spin precession can be driven either by the electric-field-tunable Rashba field\cite{koo_science09} $B^{\text{SO}}$ or by an external magnetic field\cite{crooker_science05,appelbaum_nature07,lou_np07} $B_{\text{ext}}$. Here, the device allows us to combine these two controls. The Larmor frequency for a combined field $B^{\text{SO}}+B_{\text{ext}}$ is given by $\omega_L = (2\alpha k_x - g \mu_{\text{B}} B_{\text{ext}}) / \hbar$, which determines the spin precession angle\cite{datta_apl90,serra_prb05} (Supplementary Section 3):
\vspace*{3mm}\begin{equation}
\theta = 2 m^* \alpha L /\hbar^2 - g \mu_{\text{B}} B_{\text{ext}} m^* L/ k_x \hbar^2,
\label{eq1}
\vspace*{3mm}\end{equation}
where $\alpha$ parameterizes the strength of Rashba SO coupling in the 2DEG channel, $g$ is the Land\'{e} $g$-factor, $\mu_{\text{B}}$ is the Bohr magneton, $m^*$ is the electron effective mass, and $L$ is the length between the QPC injector and detector.

Figure~4a maps the spin precession angle, manifested in the voltage oscillation, as a function of $V_{\text{M}}$ (which controls $B^{\text{SO}}_{\text{2D}}$ and thus $\alpha$ in Eq.~1) and $B_{\text{ext}}$. The external field $B_{\text{ext}}$ was applied parallel to $B^{\text{SO}}_{\text{2D}}$, both along the $y$ axis. The experimental results reveal the interplay of the electric and magnetic fields on spin precession, showing voltage oscillations along both $V_{\text{M}}$ and $B_{\text{ext}}$ axes. The dashed lines simulate the shift in the peak positions of the voltage oscillation under this interplay using Eq.~1, with the parameters $L = 2$~$\mu$m, $m^*=0.04m_e$ (where $m_e$ is the free electron mass), $k_x=1.2 \times 10^{8}$~m$^{-1}$ (estimated from the carrier density), $|g| = 9$ in InGaAs\cite{simmonds_apl08}, and $\Delta \alpha (V_{\text{M}})$ (see below). A good quantitative agreement was obtained between the experimental result and the theory.

The electric contribution to the spin precession angle, $\Delta \theta (V_{\text{M}}) = 2 m^* \Delta \alpha(V_{\text{M}}) L /\hbar^2$, and consequently the variation of the SO coupling constant with respect to the gate voltage, $|\Delta \alpha (V_{\text{M}})|$, can be estimated with a spline fitting procedure drawn through the peak and dip positions of the voltage oscillation. The fit for the spin precession angle, which manifests itself as an oscillatory voltage with a constant amplitude, and the estimated gate-dependent variation of the SO coupling constant are shown in Fig.~4b \& c. Although the geometry of the device prevents us from directly measuring the local variation of $\alpha$ under the middle gate, the relation obtained through the fit is consistent with previous work using Shubnikov-de Haas measurements\cite{nitta_prl97,koo_science09}.

A quasi-1D spin FET is anticipated to have better performance than its 2D alternatives because the current modulation due to spin precession in 2D transport is expected to be washed out by the spread of precession angles\cite{datta_apl90,sugahara_ieee10}. This is because carriers with different injection angles travel different distances between the source and drain electrodes, thereby gaining a variety of spin precession angles when they reach the drain. The QPCs---in addition to providing spin selection with nearly $100\%$ efficiency and allowing only ballistic transport electrons to be collected (to sidestep the obstacles of low injection efficiency and spin relaxation)---define a quasi-1D path between the injector and detector to eliminate the phase spread, which results in a large oscillating signal modulation in the spin FET. On the basis of device functionality and application aspects, this all-semiconductor and all-electric spin FET offers a viable route for spin information processing.

\section*{Methods Summary}

The devices were fabricated on an In$_{0.75}$Ga$_{0.25}$As/In$_{0.75}$Al$_{0.25}$As modulation-doped heterostructure (Supplementary Section 4). In reverse order of growth, the layer structure is as follows: 2 nm In$_{0.75}$Ga$_{0.25}$As (cap); 45 nm In$_{0.75}$Al$_{0.25}$As; 15 nm In$_{0.75}$Al$_{0.25}$As (Si doped); 60 nm In$_{0.75}$Al$_{0.25}$As (spacer); 30 nm In$_{0.75}$Ga$_{0.25}$As (quantum well); and 250 nm In$_{0.75}$Al$_{0.25}$As. The low-temperature carrier density and mobility of the 2DEG were measured to be $2.3\times10^{11}$ cm$^{-2}$ and $2.43\times10^5$ cm$^2$V$^{-1}$s$^{-1}$, respectively, giving a mean free path for momentum relaxation of $1.92$~$\mu$m. An insulating layer (27 nm) of SiO$_2$ was deposited on the surface of the wafer by plasma-enhanced chemical vapor deposition (PECVD). Following this, optically-defined Ti/Au surface gates were fabricated using standard optical lithography, to form bond pads. The surface gates with fine features were defined using electron-beam lithography. Measurements were performed in a dilution refrigerator, in which the devices were cooled down with a $0.3$~V bias on the surface gates to suppress random telegraph noise.

Figure 1b shows the scanning electron micrograph and circuit schematic of the spin FET device. In order to measure the conductances of both QPCs and the spin FET switching signal simultaneously, lock-in measurements were performed by applying two independent sources of (i) an a.c. voltage bias $V_{\text{exc}} = 40$~$\mu$V at $91$~Hz to the QPC injector and (ii) an a.c. current bias $I_{\text{exc}} = 1$~nA at $217$~Hz to the QPC detector. Since the QPC detector was configured as a voltage probe, a voltage develops across the QPC detector $V_{\text{QPC,d}}=I_{\text{QPC,d}} / G_{\text{QPC,d}}$ in response to the $91$~Hz a.c. current injected ballistically into and through the detector: $I_{\text{QPC,d}} = \kappa I_{\text{QPC,i}} T_{\text{QPC,d}}$, where $\kappa$ accounts for the transmission losses during transport in the semiconductor 2D channel (e.g. due to scattering; $0 < \kappa <1$), $I_{\text{QPC,i}}$ is the current emitted from the QPC injector, and $T_{\text{QPC,d}}$ is the spin-dependent transmission of the QPC detector. For clarity, the detector voltage presented here was normalized for a constant current from the injector $I_{\text{QPC,i}}=1$~nA.

\section*{Acknowledgements}
We thank C.-W. Chang, C.-C. Cheng, M. Fletcher, S. N. Holmes, C.-T. Liang, S.-T.~Lo and J.~R.~Petta for discussion and/or technical assistance on device fabrication and measurements. This work was supported by the Ministry of Science and Technology (Taiwan), the Headquarters of University Advancement at the National Cheng Kung University, and the Engineering and Physical Sciences Research Council (UK).

\section*{Author Contributions}
P.C. and S.-C.H. performed the measurements and analysed the data, in which T.-M.C. participated. L.W.S. fabricated the devices with contributions from F.S, M.P., and T.-M.C; I.F., H.E.B., and D.A.R. provided wafers; J.P.G. and G.A.C.J. performed electron-beam lithography; C.H.C. and J.C.F. contributed some measurements. T.-M.C. wrote the paper with input from S.-C.H., L.W.S., F.S., and M.P.; T.-M.C. designed and coordinated the project.



\end{document}